\pdfoutput=1

\documentclass[11pt,twoside,a4paper,cmspaper,final,collab]{cms-tdr}
\def\svnVersion{412506}\def\cmsCernNoTag{CERN-EP/2016-289}\def\cmsCernDate{\today}\def\cmsMessage{Published in the Journal of High Energy Physics as \href{http://dx.doi.org/10.1007/JHEP06(2017)106}{\doi{10.1007/JHEP06(2017)106.}}}

\begin{document}\cmsNoteHeader{SMP-14-011}

\hyphenation{had-ron-i-za-tion}
\hyphenation{cal-or-i-me-ter}
\hyphenation{de-vices}
\RCS$Revision: 405653 $
\RCS$HeadURL: svn+ssh://svn.cern.ch/reps/tdr2/papers/SMP-14-011/trunk/SMP-14-011.tex $
\newlength\cmsFigWidth
\ifthenelse{\boolean{cms@external}}{\setlength\cmsFigWidth{0.85\columnwidth}}{\setlength\cmsFigWidth{0.4\textwidth}}
\ifthenelse{\boolean{cms@external}}{\providecommand{\cmsLeft}{top}}{\providecommand{\cmsLeft}{left}}
\ifthenelse{\boolean{cms@external}}{\providecommand{\cmsRight}{bottom}}{\providecommand{\cmsRight}{right}}
\cmsNoteHeader{SMP-14-011}

\title{Measurement of electroweak-induced production of $\PW\gamma$ with two jets in pp collisions at $\sqrt{s}=8\TeV$ and constraints on anomalous quartic gauge couplings}

\date{\today}

\abstract{
A measurement of electroweak-induced production of $\PW\gamma$ and two jets is performed, where the W boson decays leptonically.
The data used in the analysis correspond to an integrated luminosity of 19.7\fbinv collected by the CMS
 experiment in $\sqrt{s}=8\TeV$ proton-proton collisions produced at the LHC\@.
Candidate events are selected with exactly one muon or electron, missing transverse momentum, one photon, and two
 jets with large rapidity separation.
An excess over the hypothesis of the standard model without electroweak production of $\PW\gamma$ with two jets
is observed with a significance of 2.7 standard deviations.
The cross section measured in the fiducial region is $10.8 \pm 4.1 \stat \pm 3.4 \syst \pm 0.3 \lum \unit{fb}$,
which is consistent with the standard model electroweak prediction.
The total cross section for $\PW\gamma$ in association with two jets in the same fiducial region is measured to be
$23.2 \pm 4.3 \stat  \pm 1.7 \syst \pm 0.6 \lum \unit{fb}$,
which is consistent with the standard model prediction from the
combination of electroweak- and quantum chromodynamics-induced processes.
No deviations are observed from the standard model predictions and experimental limits on anomalous quartic gauge
couplings $f_{M,0-7}/\Lambda^4$, $f_{T,0-2}/\Lambda^4$, and $f_{T,5-7}/\Lambda^4$ are set at 95\% confidence level.
}

\hypersetup{%
pdfauthor={CMS Collaboration},%
pdftitle={Measurement of electroweak-induced production of W gamma with two jets in pp collisions at sqrt(s)=8 TeV and constraints on anomalous quartic gauge couplings},%
pdfsubject={CMS},%
pdfkeywords={CMS, physics, aGC}}

\maketitle

\clearpage{}
\section{Introduction}
\label{sec:intro}

In the past few decades the standard model (SM) of particle physics has achieved great success through various stringent
tests and the discovery of all its predicted particles, including the recently observed Higgs
boson~\cite{Aad:2012tfa,Chatrchyan:2012ufa,CMSHiggsDiscoveryLong,Aad:2015zhl}. Additionally, the non-Abelian nature of gauge interactions was tested by the
measurements of diboson production (e.g., Refs.~\cite{Khachatryan:2014dia,CMSdiboson,Chatrchyan:2012bd,ATLAS:2012mec,Khachatryan:2015pba,Khachatryan:2015sga,Aad:2012twa,Chatrchyan:2013fya,Aad:2013izg}).
The CERN LHC allows
the measurement of many novel processes predicted by the SM,
especially those that involve pure electroweak (EW) interactions with relatively small cross sections compared
with QCD-induced production of EW final states.
Typical examples include triple gauge boson production~\cite{Chatrchyan:2014bza} and vector boson
scattering (VBS) or vector boson fusion (VBF)
processes~\cite{Aaboud:2016dkv,Aad:2014zda,Khachatryan:2014sta,Chatrchyan:2013jya,Khachatryan:2016mud,Aad:2014dta,Khachatryan:2014dea,Khachatryan:2016qkk}.

The VBS processes have some features that can be exploited to better understand the SM in novel phase spaces and to probe new physics or constrain anomalous gauge couplings.
For example, phenomenological studies of the EW production of W and Z bosons in association with two jets
that exploit the large rapidity gaps between the two jets~\cite{Rainwater:1996ud,Khoze:2002fa}.
Also, the VBF process was studied using the Higgs boson production and decay in Ref.~\cite{Rainwater:1998kj,Plehn:1999xi,Rainwater:1999sd,Kauer:2000hi}.
Furthermore, the EW production of Z bosons, $\Z\gamma$, Z$\gamma\gamma$, and same-sign W boson pairs in association with two jets has recently been measured at the LHC~\cite{Chatrchyan:2013jya,Aad:2014dta,Khachatryan:2014dea,Aad:2016sau,Aad:2014zda,Khachatryan:2014sta}.
Moreover, both the ATLAS and the CMS experiments found evidence for exclusive
$\gamma \gamma$ to $\PWp \PWm$ production~\cite{Aaboud:2016dkv,Khachatryan:2016mud},
and the ATLAS experiment found evidence for $\PW\gamma\gamma$ triple boson production~\cite{Aad:2015uqa}.
All the results are in good agreement with the SM predictions.

\begin{figure}[h!]{
\centering
    \includegraphics[width=0.31\textwidth]{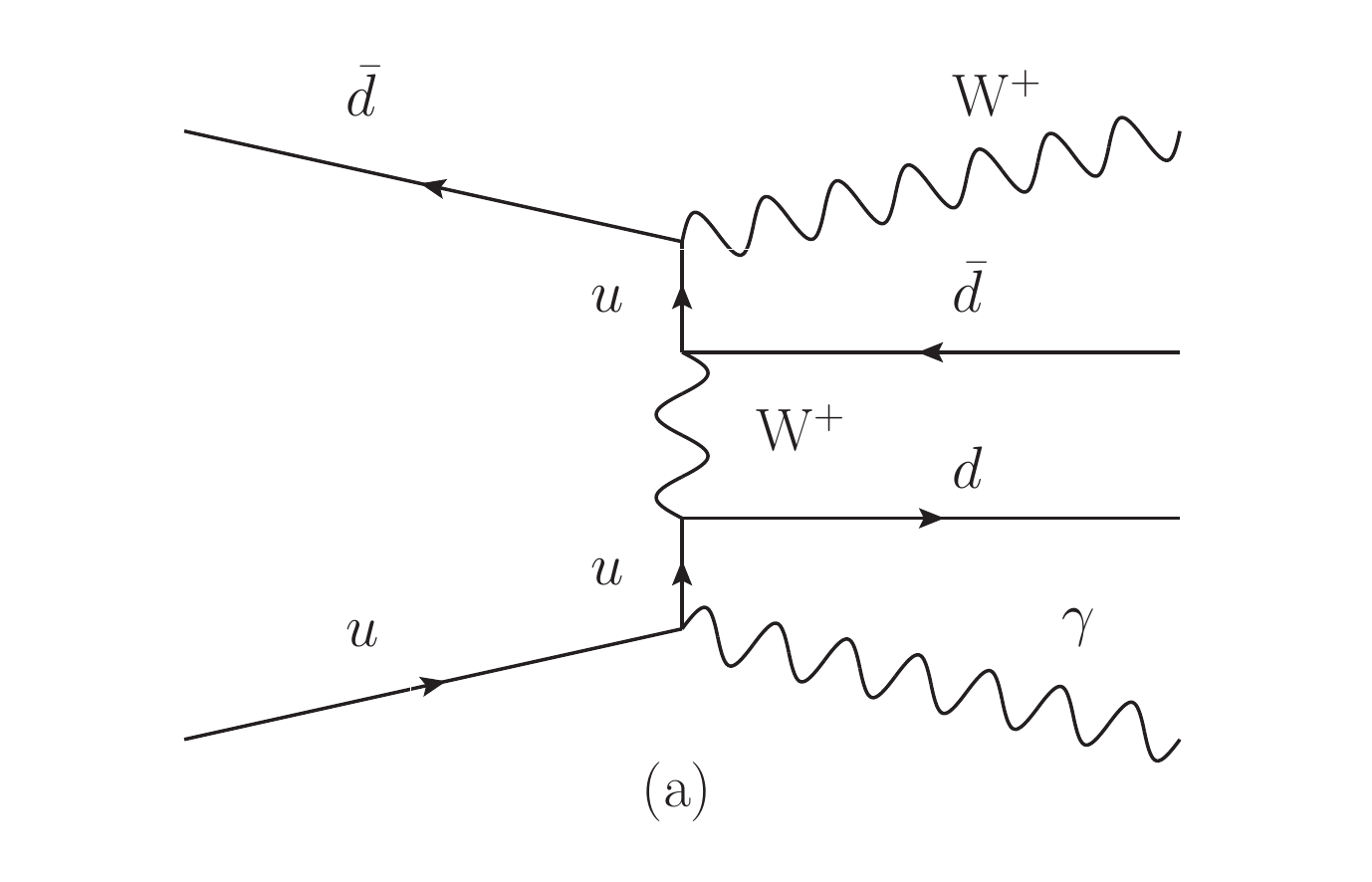}
    \includegraphics[width=0.31\textwidth]{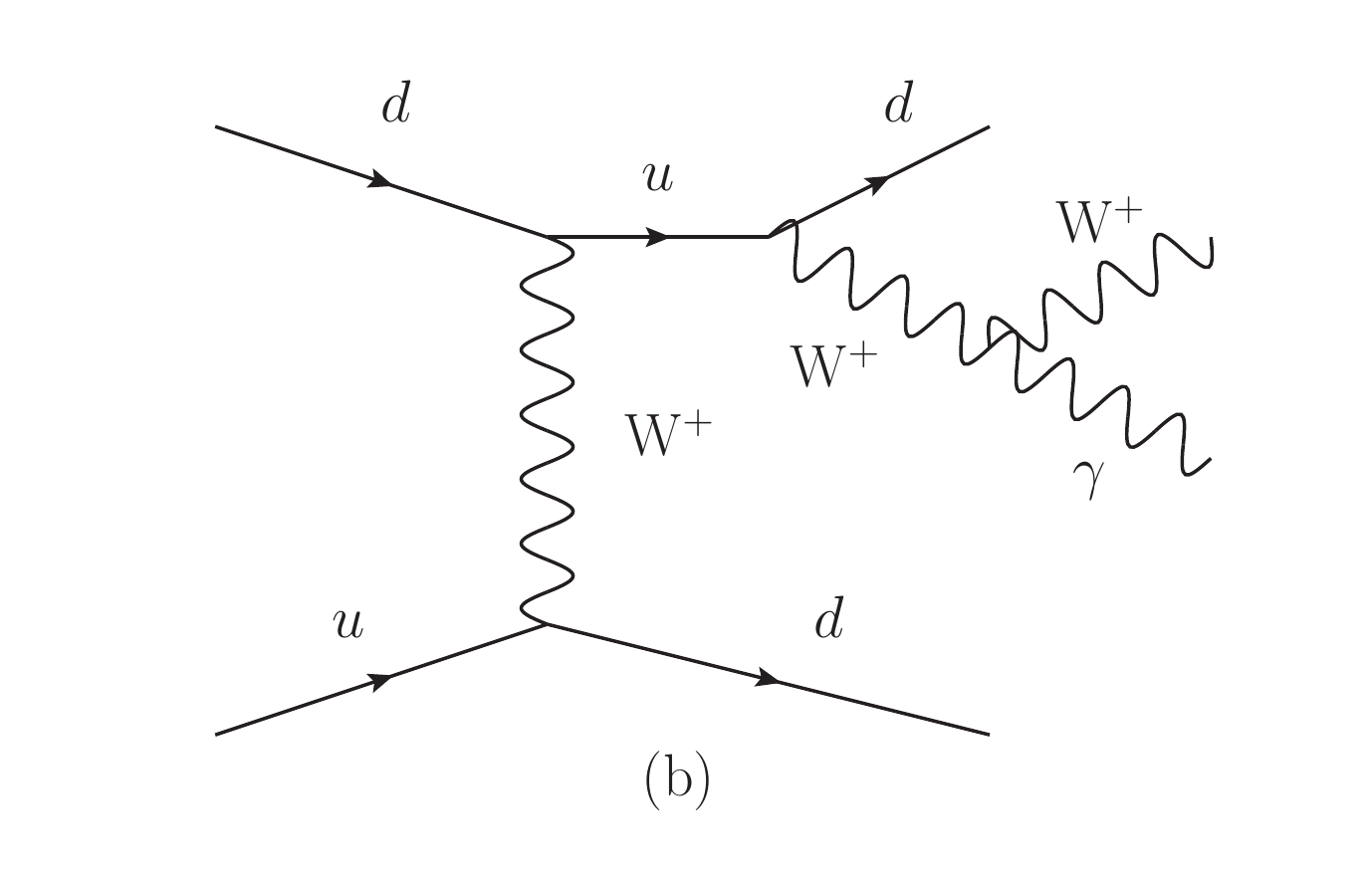}
    \includegraphics[width=0.31\textwidth]{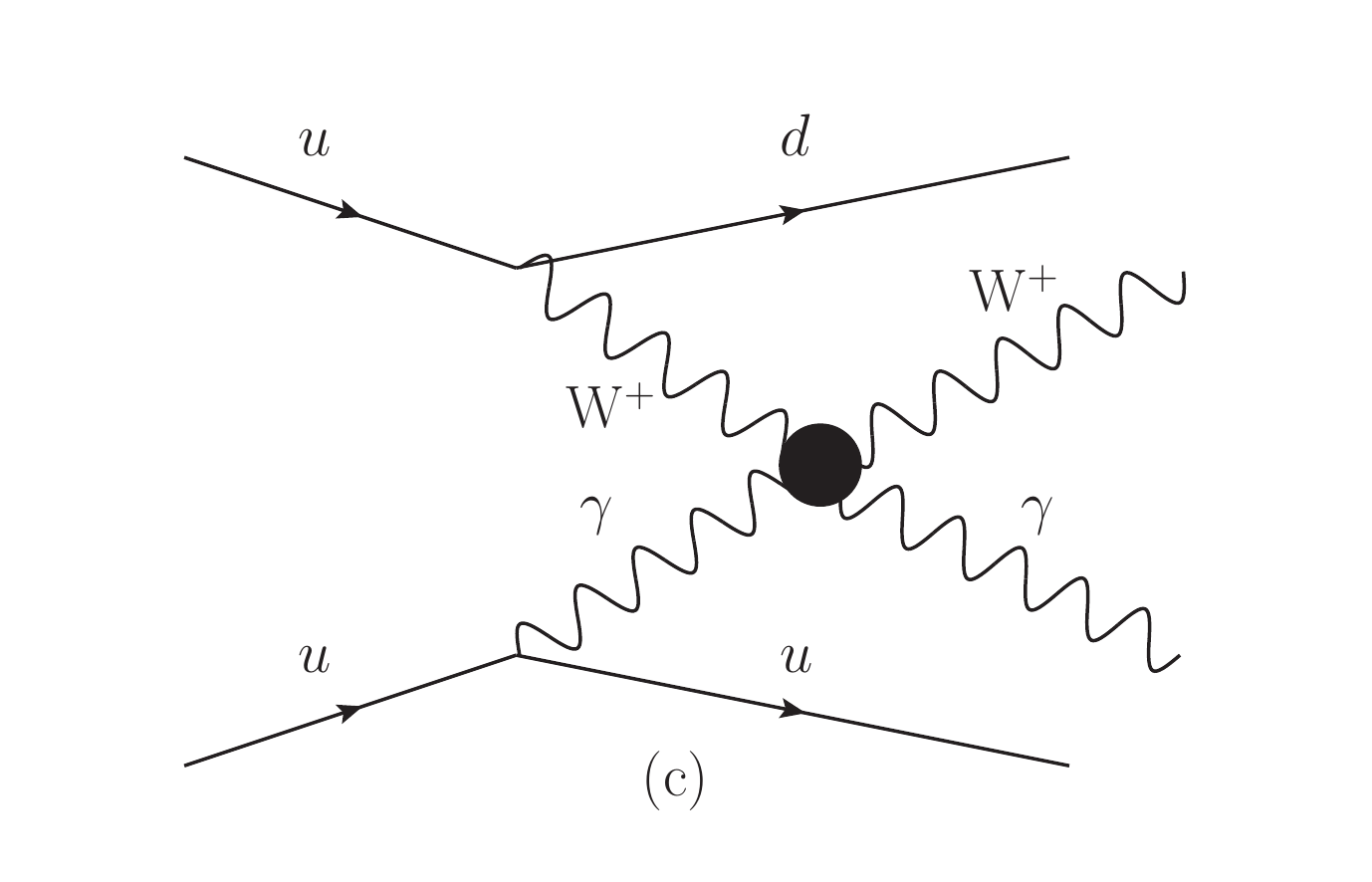}
\caption{\label{fig:wa_feynman} Representative diagrams for EW $\PW\gamma$+2\,jets production at the LHC corresponding to (a) bremsstrahlung, (b) bremsstrahlung with triple gauge coupling, and (c) VBS with quartic coupling. }}
\end{figure}

In this analysis, we search for EW-induced $\PW\gamma$ production in association with two jets~\cite{Campanario:2013eta}
(EW $\PW\gamma$+2\,jets) in the W boson leptonic decay channel ($\PW\to \ell\nu$, $\ell=\Pe,\mu$).
This process is expected to have one of the largest cross sections of all the VBS processes and thus is expected to be one of the first VBS processes observable at a hadron collider.
As shown in Fig.~\ref{fig:wa_feynman}, $\PW\gamma$ production includes several different
classes of diagrams: bremsstrahlung of one or two vector bosons and the more interesting VBS EW processes such as in
Fig.~\ref{fig:wa_feynman}c.
The cross sections of EW-induced only and EW+QCD total $\PW\gamma$ processes are measured in a VBS-like fiducial region, where the two jets have a large separation in pseudorapidity.
The signal structure of the weak boson scattering events makes VBS processes a good probe of quartic gauge boson couplings.
Instead of measuring the SM gauge couplings, which are completely fixed by the SM $SU(2)_L \otimes U(1)_Y$ gauge symmetry, we keep the SM gauge symmetry while setting limits on a set of higher dimensional anomalous quartic gauge couplings (aQGCs). More details of the aQGC parameterization can be found in Appendix~\ref{sec:aQGCth}.

The production of $\PW\gamma$+2\,jets at the LHC has two major contributions at leading order (LO)
in addition to the EW signal process described above: QCD and triple gauge boson WV$\gamma$ processes, with $\text{V} = \PW$ or $\Z$ decaying into a quark-antiquark pair.
Because these processes can have the same set of initial and final states, these three contributions interfere. One can suppress this interference by choosing an appropriate phase space for the measurements.
The WV$\gamma$ events reside mainly in the $\PW$ or $\Z$ boson mass window; we require $m_{jj} > 200\GeV$ to eliminate most of this contribution. The EW $\PW\gamma$+2\,jets events favor a larger $m_{jj}$ region than the QCD $\PW\gamma$+2\,jets events do. Calculations using the \MADGRAPH program show the interference decreases with increasing $m_{jj}$ and $|\Delta\eta(j1,j2)|$, and can change from constructive to destructive at ${\sim}1\TeV$ in $m_{jj}$ depending on the choice of renormalization and factorization scales. In the analysis we consider the phase space region with $m_{jj} > 700\GeV$ and $\abs{\Delta\eta(j1,j2)} > 2.4$ to suppress the interference. The interference effect in the fiducial region is estimated to be 4.6\% of the total $\PW\gamma$+2\,jets cross section.

In addition to the main background from QCD $\PW\gamma$+2\,jets production~\cite{Campanario:2014dpa}, other backgrounds include
(1) jets misidentified as photons or electrons, (2) WV$\gamma$ events with hadronically decaying V bosons
($\PW/\Z \to jj$) and a photon from initial- or final-state radiation, (3) contributions from top quark pairs with a radiated photon,
and (4) single top quark
events with a radiated photon. The selection criteria are designed to reduce the collective sum of these
backgrounds.
In the case of nonzero anomalous couplings, the EW contribution can be greatly enhanced, especially in the high-energy tails of some kinematic
distributions; therefore, we require the photon and W boson to have large transverse momenta to obtain better sensitivity.

The paper is organized as follows:
Section~\ref{sec:CMSdet} describes the CMS detector.
Section~\ref{sec:datamc} presents the Monte Carlo event
simulation and data sample and Section~\ref{sec:recosel} describes the event reconstruction and selection.
In Section~\ref{sec:BackgroundModeling}, methods of background modeling are explained.
Systematic uncertainties considered in the analysis are discussed subsequently in
 Section~\ref{sec:Syst}.
Results of the search for the EW signal and the measured EW and EW+QCD cross sections in the fiducial region are reported in Section~\ref{sec:xsec}.
Results on anomalous couplings using the W boson transverse momentum distribution are given in Section~\ref{sec:limits_pT}.
Finally, Section~\ref{sec:summary} summarizes the results.

\section{The CMS detector }
\label{sec:CMSdet}

The central feature of the CMS apparatus is a superconducting solenoid of 6\unit{m} internal diameter
and 13\unit{m} length, providing a magnetic field of 3.8\unit{T}. Within the solenoid volume are a silicon
pixel and strip tracker, a lead tungstate crystal electromagnetic calorimeter (ECAL), and a brass and scintillator hadron
calorimeter (HCAL). Muons are reconstructed in gas-ionization detectors embedded in the steel flux-return yoke outside the
solenoid. Extensive forward calorimetry complements the coverage provided by the barrel and endcap detectors.

The tracking system consists of 1440 silicon pixel and 15~148 silicon strip detector modules and
 covers the pseudorapidity range $\abs{\eta} < 2.5$, providing a transverse momentum \pt resolution of about 1.5\% at $100\GeV$.
The electromagnetic calorimeter consists of 75\,848 lead tungstate crystals, which provide coverage in $\abs{ \eta } < 1.48 $ in the barrel region (EB) and $1.48 < \abs{ \eta } < 3.00$ in the two endcap regions (EE). A preshower detector consisting of two planes of silicon sensors interleaved with three radiation lengths of lead is located in front of the EE. Photons are identified as ECAL energy clusters not linked to the extrapolation of any charged particle trajectory to the ECAL. These energy clusters are merged to form superclusters that are five crystals wide in $\eta$, centered around the most energetic crystal, and have a variable width in the azimuthal angle $\phi$.
The HCAL consists of a set of sampling calorimeters that utilize alternating layers of brass as absorber and plastic scintillator as active material. It provides coverage for $\abs{\eta} < 3.0$.
Combined with the forward calorimeter modules, the coverage of hadronic jets is extended to $\abs{\eta} < 5.0$.
 The energy of charged hadrons is determined from a combination of the track momentum and the corresponding ECAL and HCAL energies, corrected for the combined response function of the calorimeters. The energy of neutral hadrons is obtained from the corresponding corrected ECAL and HCAL energies.
The muon system includes barrel drift tubes covering the range $\abs{\eta} < 1.2$, endcap cathode strip chambers ($0.9 < \abs{\eta} < 2.5$),
and resistive-plate chambers ($\abs{\eta} < 1.6$)~\cite{muReco}.
The CMS detector is nearly hermetic, allowing for measurements of the missing
transverse momentum vector \ptvecmiss, which is defined as the projection on the plane perpendicular to the beams of the negative vector sum of the momenta of all reconstructed particles in an event.

The first level of the CMS trigger system, composed of custom hardware processors, uses information from the calorimeters and muon detectors to select the events of interest in a fixed time interval of less than 4\mus. The high-level trigger processor farm further decreases the event rate from around 100\unit{kHz} to
less than 1\unit{kHz}, before data storage.

A more detailed description of the CMS detector, together with a definition of the coordinate
system used and the relevant kinematic variables, can be found in Ref.~\cite{Chatrchyan:2008zzk}.

\section{ Data and simulated samples }
\label{sec:datamc}

The analysis uses a data sample of proton-proton collisions collected at $\sqrt{s}=8\TeV$ by the CMS detector in 2012 that
corresponds to an integrated luminosity of $19.7 \pm 0.5\fbinv$~\cite{lumiPAS}.

{\tolerance=1400
The analysis makes use of several simulated event samples based on
Monte Carlo (MC).
The EW $\PW(\to \ell\nu)\gamma$+2\,jets process and the $\ttbar\gamma$ background process are generated using \MADGRAPH 5.1.3.22~\cite{MadGraph}.
Samples with aQGCs are obtained using the multi-weight method with the \MADGRAPH 5.2.1.1 generator~\cite{Alwall:2014hca}.
The MC samples for QCD $\PW(\to \ell\nu)/\Z(\to \ell\ell)\gamma$+0,1,2,3 jets are also generated with the \MADGRAPH 5.2.1.1 generator,
using the MLM matching
method~\cite{Catani:2001cc,Alwall:2008qv,Alwall:2014hca,Alwall:2007fs} with
a matrix element/parton shower (ME-PS) matching scale of 10\GeV~\cite{Hoche:2006ph}.
For all samples generated with \MADGRAPH, the CTEQ6L1 parton distribution function (PDF) set~\cite{Pumplin:2002vw} is used,
and the renormalization and factorization scales are set to
$\sqrt{\smash[b]{M_{\PW/\Z}^2 + (\pt^{\PW/\Z})^2 + (\pt^{\gamma})^2 + \sum (\pt^{j})^2}}$.
The single top quark production processes are generated with the {\POWHEG} (v1.0, r1380)~\cite{Alioli:2009je,Re:2010bp} generator, using
the CTEQ6M PDF set~\cite{Pumplin:2002vw,Nadolsky:2008zw}.
The diboson samples (WW, WZ, ZZ), with one of the bosons decaying leptonically and the other decaying hadronically, are generated with {\PYTHIA}~6.422~\cite{pythia} and the CTEQ6L1 PDF set.
The final-state leptons considered are $\Pe,\Pgm,$ and $\tau$, where the $\tau$ lepton decay is handled with
{\TAUOLA}~\cite{Jadach:212328}. The {\PYTHIA} 6.426~\cite{pythia} program is used to simulate parton showers and
hadronization, with the parameters of the underlying event set to the Z2* tune~\cite{PythiaTuneZ2,Chatrchyan:2013gfi}.
\par}

For all MC samples, a \GEANTfour-based simulation~\cite{GEANT4} of the CMS detector is used and the hard-interaction
collision is overlaid with a number of simulated minimum-bias collisions. The resulting events are weighted to reproduce the
data distribution of the number of inelastic collisions per bunch crossing (pileup). These simulated events are
reconstructed and analyzed using the same algorithms as for data. The differences in lepton and photon
reconstruction and identification (ID) efficiencies observed between data and simulated events are subsequently corrected with
scale factors~\cite{Khachatryan:2015hwa,Khachatryan:2015iwa}.

To improve the precision of the predicted cross section for the signal model,
the NLO QCD correction is included with the EW signal process through an
NLO/LO cross section $K$ factor of 1.02, determined by using {\sc vbfnlo}~\cite{Baglio:2014uba,Arnold:2011wj,Arnold:2008rz,Campanario:2013eta,Campanario:2014dpa}.
For QCD $\PW\gamma$+2\,jets production, the $K$ factor is 0.93 and is only applied for the measurement of the EW+QCD cross section, fixing the ratio between EW and QCD components.

\section{Event reconstruction and selection}
\label{sec:recosel}

An EW-induced $\PW\gamma$+2\,jets event is expected to have exactly one
lepton (muon or electron), a photon, two jets with large rapidity separation, and large $\abs{\ptvecmiss}$.

A complete reconstruction of the individual particles emerging from each collision event is obtained via a particle-flow (PF) technique, which uses the information from all CMS subdetectors to identify and reconstruct individual particles~\cite{PFT-09-001,CMS:2010byl}. The particles are classified into mutually exclusive categories: charged hadrons, neutral hadrons, photons, muons, and electrons.

The events are selected by using single-lepton triggers with \pt thresholds of
24\GeV for muons and 27\GeV for electrons.
The overall trigger efficiency is 90\% (94\%) for the electron (muon) data, with a small dependence on \pt and $\eta$.
Charged-particle tracks are required to originate from the event
primary vertex, defined as the reconstructed vertex within
24 cm (2 cm) of the center of the detector in the direction along
(perpendicular to) the beam axis that has the highest value
of $\pt^2$ summed over the associated charged-particle tracks.

The events are also required to have either one muon or one electron;
events with additional charged leptons are excluded.
The muon candidates are reconstructed with information from both the silicon tracker and from the muon detector by means of a
global fit~\cite{muReco}.
They are required to satisfy a requirement on the PF-based relative isolation, which is defined as the ratio of the $\pt$ sum of all other PF candidates reconstructed in a cone $\Delta R = \sqrt{\smash[b]{(\Delta\eta)^2+(\Delta\phi)^2}} = 0.3$ $(0.4)$ around the candidate electron (muon) to the $\pt$ of the candidate, and is corrected for contributions from pileup~\cite{Khachatryan:2015hwa}.
The selection efficiency is approximately
96\%. Muons with $\pt > 25\GeV$ and $\abs{\eta} < 2.1$ are included in the analysis.
The electron candidates are reconstructed by associating a charged
particle track originating from the event primary vertex with
superclusters of energy depositions in ECAL~\cite{Khachatryan:2015hwa}.
They must also satisfy the PF-based relative isolation be smaller than 0.15.
The ID and isolation selection
efficiency is approximately 80\%.
The electron candidates are further required to satisfy $\pt > 30\GeV$ and $\abs{\eta} < 2.5$, excluding the transition region between the
ECAL barrel and endcaps, $1.44 < \abs{\eta} < 1.57$, because the reconstruction of electrons in this region has lower efficiency. To suppress the $\Z\to \Pep\Pem$ background in the electron
channel, where one electron is misidentified as a photon, a {\Z} boson mass veto of $\abs{m_{\Pe\gamma}-M_{\Z}} > 10\GeV$ is applied.

A well-identified and isolated photon is also required for the event selection~\cite{Khachatryan:2015iwa}.
Photons are
reconstructed from superclusters and are
required to satisfy a
number of criteria aimed at rejecting misidentified jets.
They have to have a small ratio of hadronic energy in the HCAL that is matched in $(\eta,\phi)$ to
the electromagnetic energy in the ECAL;
small shower shape variable $\sigma_{\eta \eta}$, which quantifies the
 lateral extension of the shower along the $\eta$ direction~\cite{Khachatryan:2015hwa};
small PF-based charged and neutral
photon isolations including pileup corrections~\cite{PFT-09-001}; and an electron-track veto to reduce electron
misidentification.
With these requirements the photon ID and isolation efficiency is about 70\%.
The resulting photon candidates are further required to satisfy $\pt^{\gamma} > 22\GeV$ and must be in the barrel region with
$\abs{\eta_{\mathrm{sc}}} < 1.44$, where $\eta_{\mathrm{sc}}$ refers to the supercluster $\eta$, corresponding to a fiducial region in the ECAL barrel excluding the outer barrel ECAL rings of crystals.

Jets are reconstructed from PF particles~\cite{PFT-09-001,CMS:2010byl} using the anti-\kt clustering algorithm~\cite{ANTIKT} with a distance parameter of 0.5.
Only charged particles with tracks originating from the primary vertex are considered for clustering.
Jets from pileup are identified and removed with a pileup jet identification algorithm~\cite{CMS-PAS-JME-13-005}, based on both vertex information and jet shape information. Jets are required to satisfy a set of loose ID criteria designed to
eliminate jets originating from noisy channels in the calorimeter~\cite{Chatrchyan:2009hy}.
Pileup collisions and the underlying event can contribute to the energy of the reconstructed jets.
A correction based on the projected area of a jet on the front face of the calorimeter is used to subtract the extra energy deposited in the jet coming from
pileup~\cite{fastjet1,Cacciari:2008gn}. Furthermore, the energy response in $\eta$ and \pt is corrected, and the energy resolution is smeared for simulated samples to
give the same response as observed~\cite{Chatrchyan:2011ds}.
An event is selected if it has at least two jets, with the leading jet $\pt > 40\GeV$, second-leading jet $\pt > 30\GeV$, and each jet within $\abs{\eta} < 4.7$.
These two jets are denoted as ``tag jets''.
To suppress the WV$\gamma$ background, $m_{jj}$ is required to be at least 200\GeV.

In addition, the event should have $\abs{\ptvecmiss} > 35\GeV$.
The reconstructed transverse mass of the leptonically decaying W boson, defined as
$M_{\mathrm{T}}=\sqrt{2\smash[b]{{\pt^{\ell}\abs{\ptvecmiss}[1-\cos(\Delta\phi_{\ell,\ptvecmiss})]}}}$, where $\Delta\phi_{\ell,\ptvecmiss}$ is the azimuthal angle
between the lepton momentum and the \ptvecmiss, is required to exceed 30\GeV~\cite{CMS:2011aa}. We reconstruct the leptonic W boson decay
by solving for the longitudinal component of the neutrino momentum and using the mass of the W boson as a constraint.
In the case of complex solutions in this reconstruction, we choose the real part of the solution, and if there are two real solutions,
we choose the solution that gives a neutrino momentum vector that is closer to the longitudinal component of the corresponding charged lepton momentum.

Mismeasurement of jet energies can generate $\abs{\ptvecmiss}$.
To eliminate events in which this mismeasurement may generate an
apparent large $\abs{\ptvecmiss}$, the azimuthal separation
between each of the tag jets
and the \ptvecmiss is required to be larger than 0.4\unit{rad}. Additionally, to suppress the top quark backgrounds, we require that the tag jets
fail a {\cPqb} tagging requirement of the combined secondary vertex algorithm~\cite{Chatrchyan:2012jua} with a misidentification rate of 1\%.

Separation between pairs of objects in the event is required:
$\Delta R_{jj}$, $\Delta R_{j \gamma}$, $\Delta R_{j \ell}$, and
$\Delta R_{\ell \gamma}> 0.5$. All the requirements described above
ensure the quality of the identified final states and comprise the baseline selections for the analysis. Table~\ref{tab:BS-SELECTION} summarizes these criteria.
\begin{table}[tb]
\centering
\topcaption{Summary of the baseline selection criteria.}
\label{tab:BS-SELECTION}
\begin{tabular}{l l}
\hline
          Single-lepton ($\Pe,\Pgm$) trigger                  & $|M_{\Pe\gamma}-M_{\Z}|>10 \GeV$ (electron channel) \\
          Lepton, photon ID and isolation        & $\pt^{j1} > 40\GeV$, $\pt^{j2} > 30\GeV$ \\
          Second lepton veto                     & $\abs{\eta^{j1}}<4.7$, $\abs{\eta^{j2}}<4.7$ \\
          Muon (electron) $\pt > 25$ (30)\GeV, $\abs{\eta}<2.1 ~(2.4)$ & $\abs{\Delta\phi_{j1,\ptvecmiss}}>0.4$, $\abs{\Delta\phi_{j2,\ptvecmiss}}>0.4\unit{rad}$ \\
          Photon $\pt^{\gamma} > 22\GeV$, $\abs{\eta}<1.44$ & b quark jet veto for tag jets \\
          W boson transverse mass $> 30\GeV$         &Dijet invariant mass $m_{jj} > 200\GeV$ \\
          $\abs{\ptvecmiss} > 35\GeV$                         & $\Delta R_{jj}$, $\Delta R_{j \gamma}$, $\Delta R_{j \ell}$, $\Delta R_{\ell \gamma}$ $>$ 0.5 \\
\hline
\end{tabular}
\end{table}

To optimize the measurement of the EW-induced $\PW\gamma$+2\,jets signal and improve the EW signal significance, we further consider selections on the following variables to
suppress backgrounds: the Zeppenfeld variable~\cite{Rainwater:1996ud}, $\abs{y_{\PW\gamma} -(y_{j1} + y_{j2})/2}$, calculated using the rapidities ($y$) of the $\PW\gamma$ system and the two jets; the azimuthal separation between
the $\PW\gamma$ system, which combines the four
momenta of the W boson and the photon, and the dijet system $|\Delta\phi_{\PW\gamma,jj}|$; the dijet invariant mass $m_{jj}$;
and the pseudorapidity separation between the tag jets $|\Delta\eta(j1,j2)|$.
These additional requirements are chosen as follows:
\begin{itemize}
\item $\abs{y_{\PW\gamma} -(y_{j1} + y_{j2})/2} < 0.6$;
\item $\abs{\Delta\phi_{\PW\gamma,jj}} > 2.6\unit{rad}$;
\item $m_{jj} >  700\GeV$;
\item $\abs{\Delta\eta(j1,j2)} >  2.4$.
\end{itemize}

\section {Background estimation}
\label{sec:BackgroundModeling}

The dominant background comes from QCD $\PW\gamma$+jets production.
It is estimated using simulation and is normalized to the number of events in data in the region $200 < m_{jj} < 400\GeV$.
The data/simulation normalization factors $0.77 \pm 0.05$ (muon channel) and $0.77 \pm 0.06$ (electron channel) are consistent with the $K$ factor of $0.93 \pm 0.27$ obtained with \textsc{vbfnlo}.
For the combined measurement of the EW+QCD cross section, the contribution
of QCD $\PW\gamma$+jets is taken directly from simulation (scaled by the $K$ factor)
since this contribution is then no longer a background.

The background from misidentified photons arises mainly from
W+jets events where one jet satisfies the photon ID criteria.
The estimation is based on events similar to the ones selected with the baseline selection described in Section~\ref{sec:recosel}, except that the photon must fail the tight photon ID and satisfy a looser ID
requirement. This selection ensures that the photon arises from a jet, but
still has kinematic properties similar to a genuine photon originating from the primary vertex.
The selected events are then normalized to the number of events satisfying the tight photon ID
and weighted with the probability of a jet to be misidentified as a photon.
The misidentification probability is calculated as a function of photon \pt in a manner similar to that described in Ref.~\cite{Vgamma}. The method uses the shapes of the $\sigma_{\eta \eta}$ and PF charged isolation distributions, which differ for genuine and misidentified photons.
The fraction of the total background in the signal region
 contributed by this source decreases
with $\pt^{\gamma}$, from a maximum of 33\% ($\pt \approx 22\GeV$) to 6\%
($\pt>135\GeV$).

The $\gamma$+jets events contribute to the background when the jet is misidentified as a muon or electron.
The contribution is found to be negligible in the muon channel,
but can be significant in the electron channel, especially in the low-$m_{jj}$ region.
A control data sample is selected, in a similar way
to that discussed in the previous paragraph, from the PF
relative isolation sideband with a very loose electron ID requirement.
Events in this control sample are then normalized to the events with signal selection and weighted with the misidentification probability for a jet to satisfy the electron selections.
This probability is determined from a three-component fit to the $\abs{\ptvecmiss}$ distribution considering the $\gamma$+jets misidentified events, QCD $\PW\gamma$+jets events, and misidentified photon events, as explained in more detail in Ref.~\cite{CMS:2011aa}.
The $\gamma$+jets background contribution in electron channel is estimated to be 7\% of the total yield for the baseline selections and negligible in the EW signal region.

Other background contributions are small and are estimated from simulation.
The contributions from top quark pair and single top quark production, each in association with a photon, are suppressed with the b quark veto and represents only 3.4\% of the total event yield in the EW signal region.
The Z($\to \ell\ell$)$\gamma$(+jets) events can contribute if one of the decayed leptons is undetected, resulting in $\abs{\ptvecmiss}$.
The predicted cross sections of the $\Z\gamma$ and WV processes decrease with increasing $m_{jj}$ and contribute about 2\% of the total SM prediction in the EW signal region.

\begin{figure}[tb]{
\centering
\includegraphics[width=0.43\textwidth]{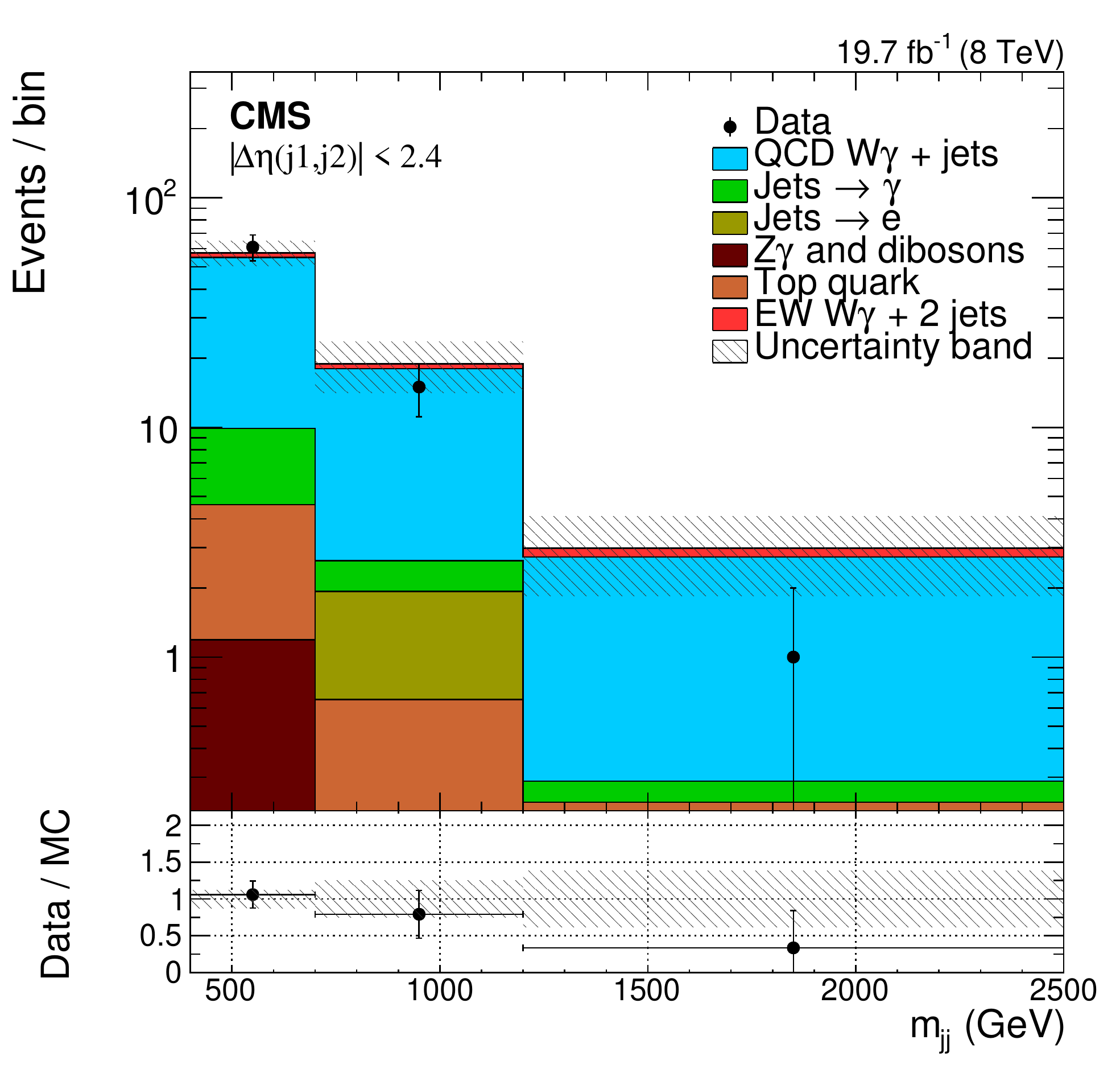}
\includegraphics[width=0.43\textwidth]{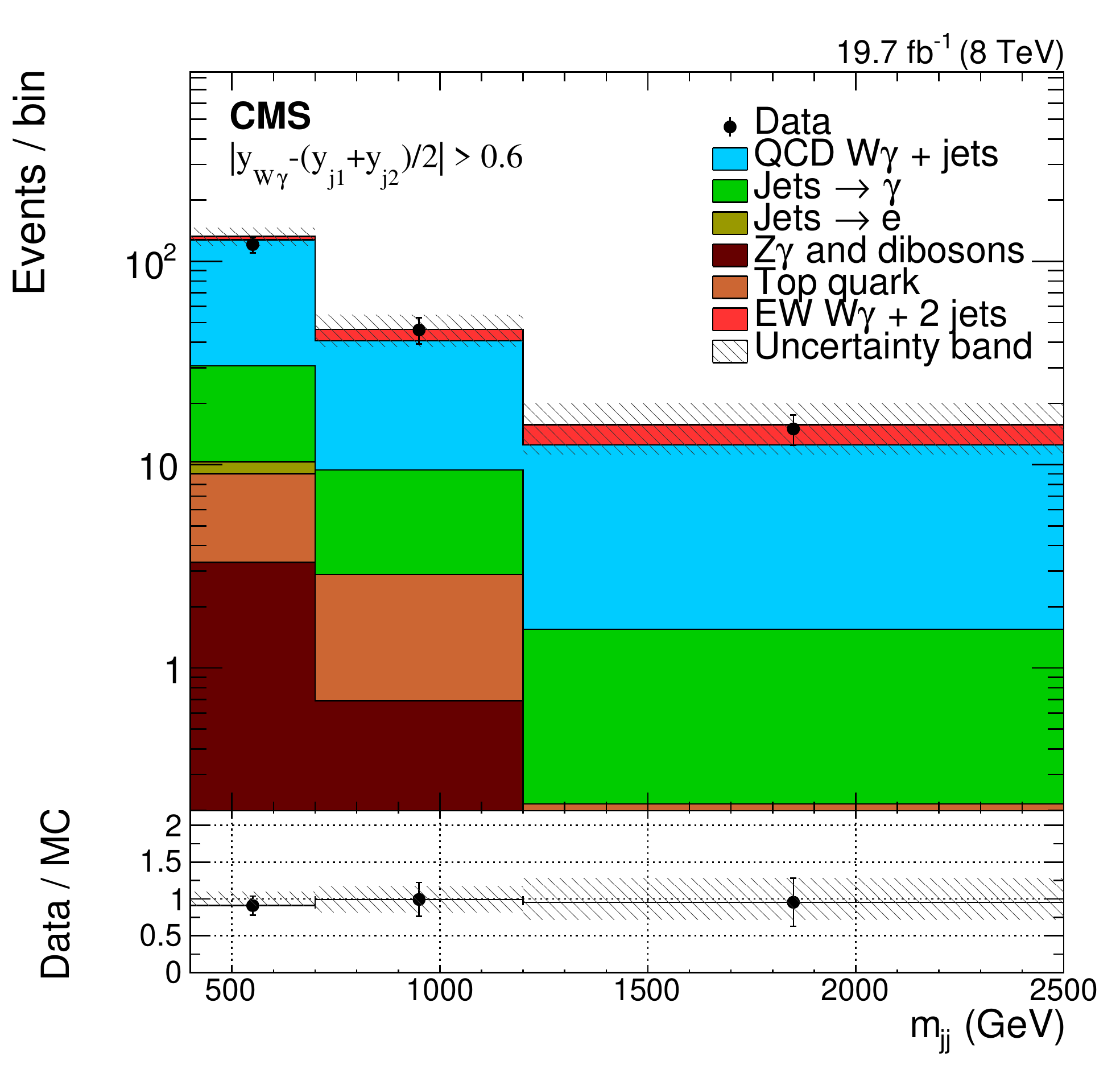}
\includegraphics[width=0.43\textwidth]{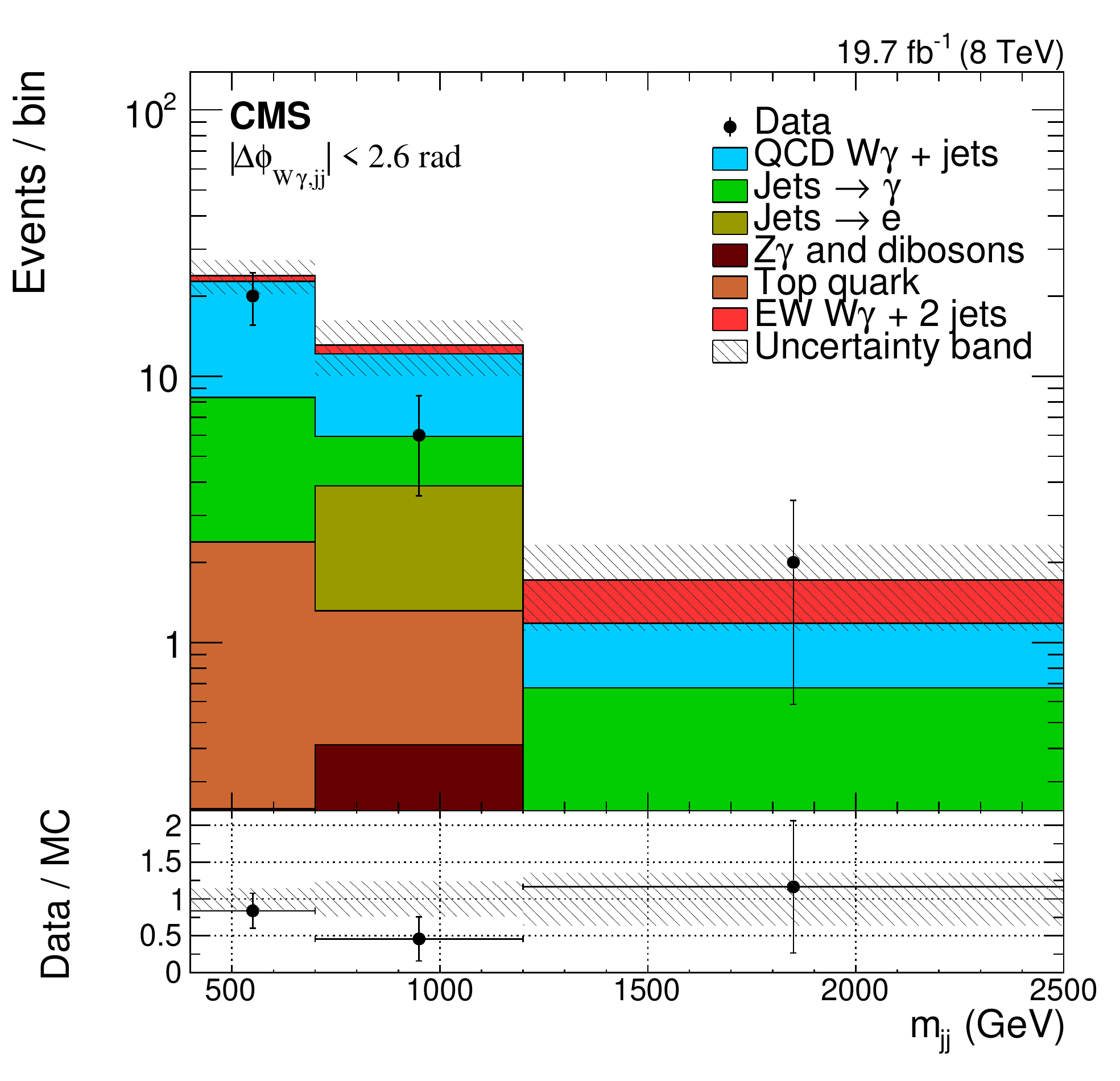}
\caption{\label{fig:contr_mjj} The $m_{jj}$ distributions in orthogonal, but signal-like, regions obtained by inverting
the signal selection criteria: $\abs{\Delta\eta(j1,j2)}<2.4$; $\abs{y_{\PW\gamma} -(y_{j1} + y_{j2})/2}>0.6$; and $|\Delta\phi_{\mathrm{W}\gamma,jj}|<2.6$\unit{rad}.
Events from electron and muon channels are combined.
Backgrounds from jets misidentified as photons ($\text{Jets} \to \gamma$) and jets misidentified as electrons ($\text{Jets} \to \Pe$) are estimated from data as described in the text.
The diboson contribution includes WV(+$\gamma$) and $\Z\gamma$(+jets) processes. The top quark contribution includes both the
$\ttbar\gamma$ and single top quark processes.
The signal contribution is shown on top of the backgrounds. The last
bin includes the overflow events.
The shaded area represents the total uncertainty in the simulation, including statistical and systematic effects.
}}
\end{figure}

Figure~\ref{fig:contr_mjj} shows three $m_{jj}$ distributions in orthogonal, but signal-like, regions obtained by inverting each of three signal selection criteria: $|\Delta\eta(j1,j2)|<2.4$; $|y_{\mathrm{W}\gamma} -(y_{j1} + y_{j2})/2|>0.6$; and $|\Delta\phi_{\mathrm{W}\gamma,jj}|<2.6$ rad.
Each of these regions is enriched in QCD production of $\PW\gamma$+jets events and, to a lesser degree, background having a jet misidentified as a photon.
They confirm our modeling of those backgrounds.

\section{Systematic uncertainties}
\label{sec:Syst}

The background rate of QCD $\PW\gamma$+jets production is measured in the low-$m_{jj}$ control region and extrapolated to the signal region.
The rate uncertainty includes the statistical uncertainty as well as the uncertainties
due to the misidentification probability of jets as photons or leptons.
This uncertainty is 6.2\% (7.1\%) for the muon (electron) channel.
In addition, when extrapolating from the control region to the signal region,
the shape dependence on theoretical parameters affects the normalization of the QCD $\PW\gamma$+jets distribution at high $m_{jj}$.
This extrapolation uncertainty is calculated by using different MC samples with matching and renormalization/factorization scales varied up and down by a factor of two.
Contributions of all the shapes are normalized in the control region and the largest absolute difference from the nominal one in the signal region is considered as the uncertainty, this is about 20\% for $m_{jj}\approx 1\TeV$.

The uncertainty on the misidentification probability of jets as electrons is estimated by considering both the $\abs{\ptvecmiss}$ fit uncertainty and shape uncertainty and is estimated to be 40\%.
There are three contributions to the uncertainties in the misidentified photon background:
the statistical uncertainty, the variation in the choice of the charged isolation sideband, and the $\sigma_{\eta \eta}$ shape
in the sample of events with objects misidentified as photons.
The combined uncertainty, calculated in $\pt^{\gamma}$ bins, increases from 13\% at $\pt^{\gamma} \approx 25\GeV$ to 54\% for $\pt^{\gamma} \approx 135\GeV$.

The uncertainty in the measured value of the
integrated luminosity is 2.6\%~\cite{lumiPAS}.
Jet energy scale and resolution uncertainties contribute via selection thresholds for the
jet \pt and $m_{jj}$. We consider the uncertainties in different intervals of $m_{jj}$, giving a
combined uncertainty varying from 12 to 31\% with increasing $m_{jj}$ in the signal region. A small
difference in $\abs{\ptvecmiss}$ resolution~\cite{Chatrchyan:2011tn} between data and simulation
affects the signal selection efficiency by less than 1\%.
The uncertainties due to the lepton trigger efficiency and reconstruction and the selection efficiencies are estimated to be 1\% and 2\%, respectively.
Photon reconstruction efficiency and energy scale uncertainties contribute to the signal selection efficiency at the 1\% level.
The uncertainty from the b jet veto procedure is 2\% in the data/simulation
efficiency correction factor~\cite{Chatrchyan:2012jua}. This
uncertainty has an effect of 8\% on the $\ttbar\gamma$ background, 23\% on the single top quark
background, and a negligible effect on the signal. The theoretical
uncertainty in the $\ttbar\gamma$ and
$\Z\gamma$+jets production cross section is 20\%~\cite{Chatrchyan:2014bza}.

The theoretical uncertainty
is evaluated with {\sc vbfnlo} by
varying the renormalization and factorization
scales, each by factors of 1/2 and 2 with the requirement that the two scales remain equal.
The envelope of all the variations is taken as the uncertainty.
The uncertainty related to the PDF is calculated using the CTEQ6.1~\cite{Stump:2003yu} PDF uncertainty sets, following the prescription of Ref.~\cite{Stump:2003yu}.
For EW $\PW\gamma$+2\,jets and possible aQGC signal yield, this uncertainty is found to be 20\% with scale variations and 2.8\% with PDF sets.
For QCD $\PW\gamma$+2\,jets, this is 29\% with scale variations and 4.2\% with PDF sets.
The theoretical uncertainties due to scale and PDF choices affect the expected $m_{jj}$ shape and introduce an uncertainty
in the cross section measured by fitting the $m_{jj}$ distribution.
In addition, they affect the signal and the selection acceptance and efficiency.
Extrapolation from the selected region to the fiducial cross section region, defined in Section~\ref{sec:xsec}, introduces an uncertainty of 1\% in the measured fiducial cross section.

\section{\texorpdfstring{EW $\PW\gamma$+2\,jets signal and cross section measurements}{EW W gamma+2 jets signal and cross section measurements}}
\label{sec:xsec}

\begin{figure}[tb]{
\centering
\includegraphics[width=0.45\textwidth]{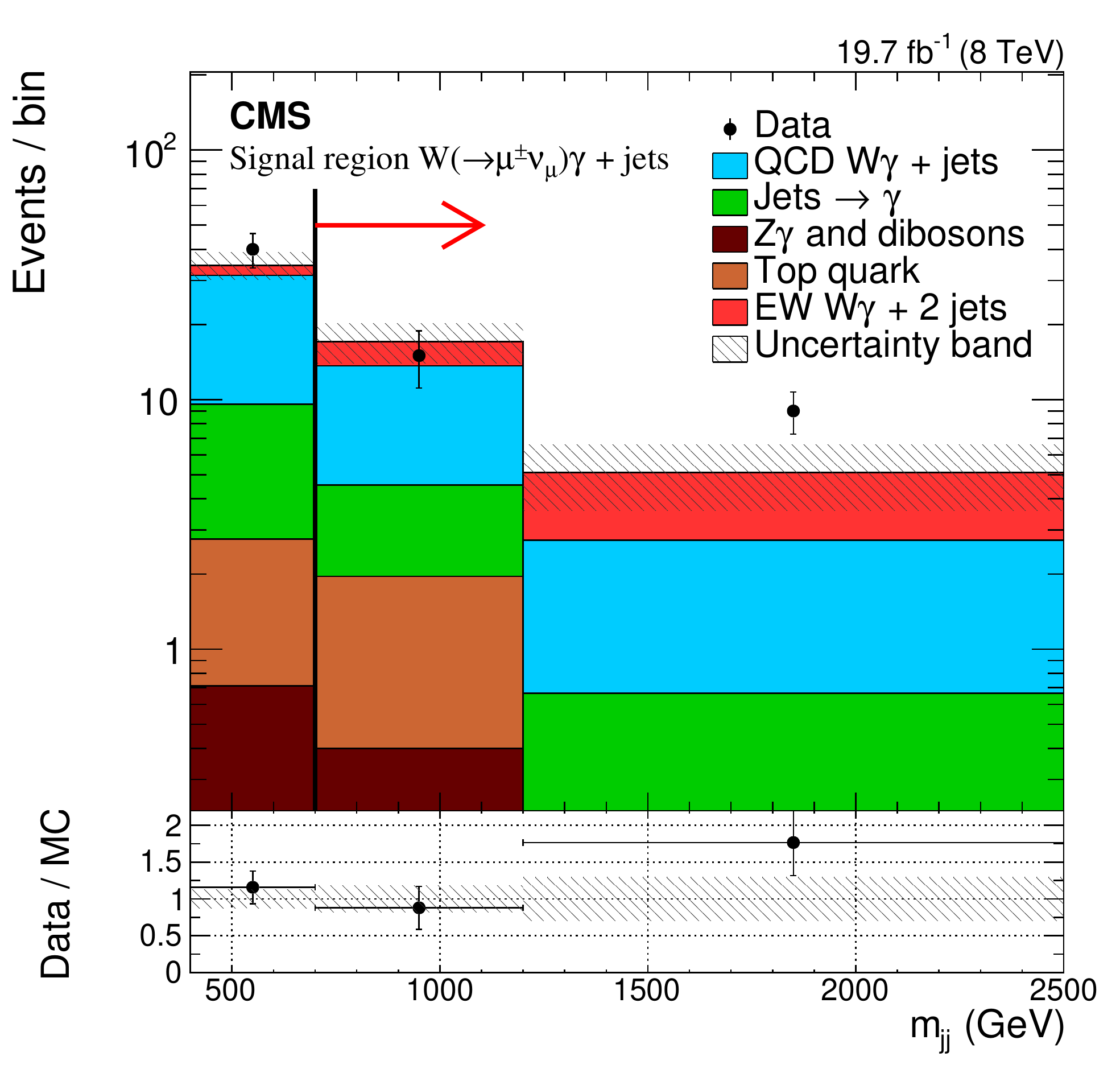}
\includegraphics[width=0.45\textwidth]{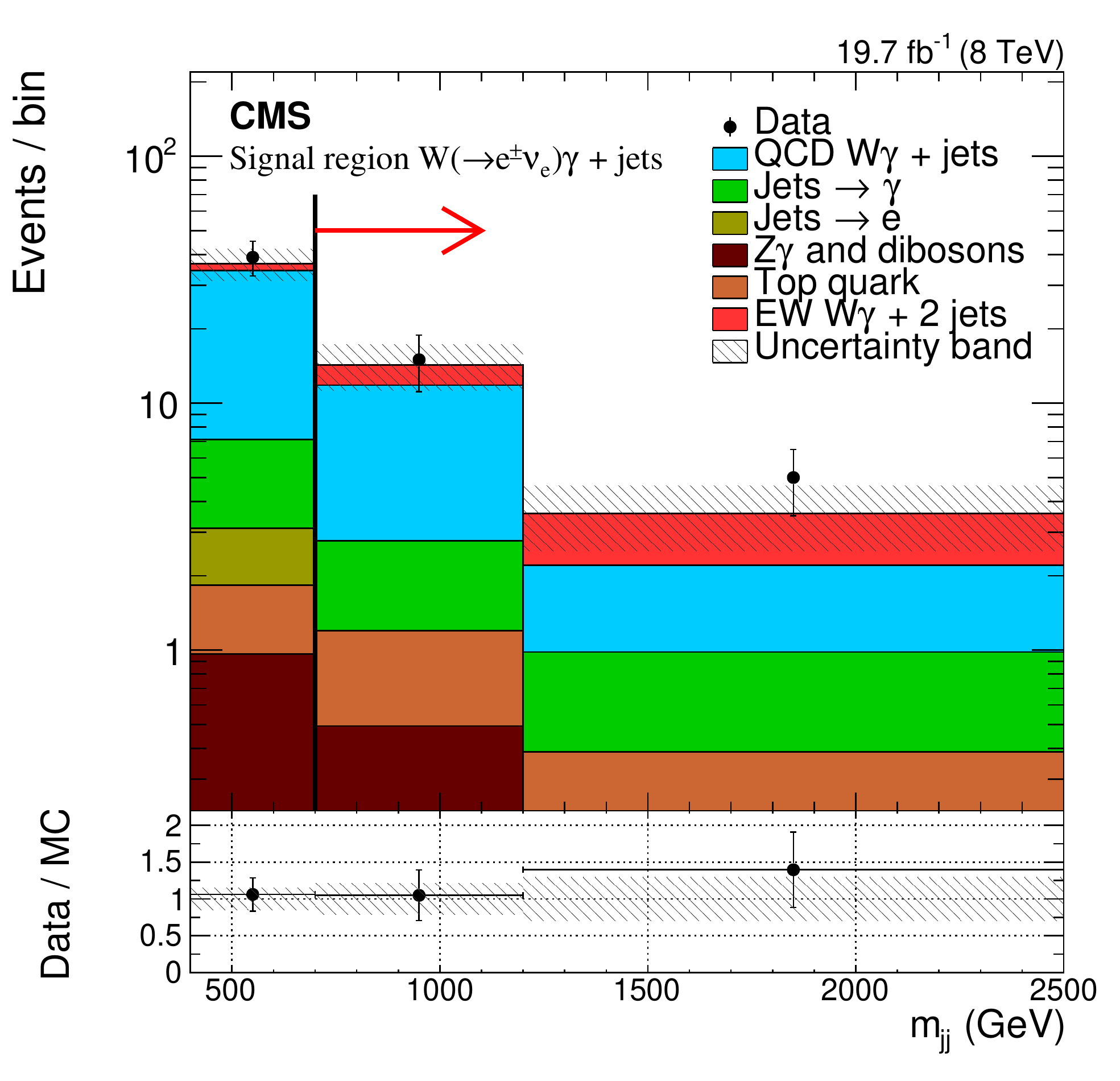}
\caption{\label{fig:mjj_plots} The $m_{jj}$ distribution in the muon (left) and electron (right) channels, in which the signal region lies above 700\GeV, indicated by the horizontal thick arrows.
Backgrounds from jets misidentified as photons (Jets $\to \gamma$) and jets misidentified as electrons (Jets $\to \Pe$) are estimated from data as described in the text.
The diboson contribution includes WV(+$\gamma$) and $\Z\gamma$(+jets) processes. The top quark contribution includes both the
$\ttbar\gamma$ and single top quark processes.
The signal contribution is shown on top of the backgrounds. The last
bin includes the overflow events.
The shaded area represents the total uncertainty in the simulation, including statistical and systematic effects.
}}
\end{figure}

A search for the SM EW $\PW\gamma$+2\,jets signal is performed based on the binned $m_{jj}$ distribution, as shown in Fig.~\ref{fig:mjj_plots}, for both the muon and electron channels,
using only the two rightmost bins corresponding to $m_{jj}>700\GeV$.
The EW- and QCD-induced $\PW\gamma$+2\,jets production is modeled at LO, neglecting interference, with NLO QCD corrections to the cross section applied through their $K$ factors.

We search for an enhancement in the rate of $\PW\gamma$+2\,jets production due to EW-induced production, treating non-$\PW\gamma$ and QCD-induced $\PW\gamma$+2\,jets production as background.
The expected signal and background yields after the selections are shown in Table~\ref{tab:evt}.

\begin{table}[tb]
\centering
  \topcaption{Number of events for each process, with combined statistical and systematic uncertainties. The total prediction
represents the sum of all the individual contributions. The W+jets background, with one jet misidentified as an electron,
is negligible in the signal region. }
  \begin{tabular}{l l l}
  \hline
  Process  & Muon channel & Electron channel  \\
  \hline
  EW-induced $\PW\gamma$+2\,jets             & $5.8   \pm 1.8$  & $3.8   \pm 1.2$  \\
  \hline
  QCD-induced $\PW\gamma$+jets                & $11.2 \pm 3.2$ & $10.3 \pm 3.2$  \\
  W+jets, 1 jet $\to \gamma$  & $3.1  \pm 0.7$  & $2.2  \pm 0.5$  \\
  MC $\ttbar\gamma$                           & $1.2  \pm 0.6$  & $0.4   \pm 0.2$  \\
  MC single top quark                         & $0.5   \pm 0.5$  & $0.6   \pm 0.4$  \\
  MC WV$\gamma$, V$\to$ two jets   & $0.3  \pm 0.2$  & $0.3   \pm 0.2$  \\
  MC $\Z\gamma$+jets                        & $0.2   \pm 0.2$  & $0.3   \pm 0.2$  \\
  \hline
  Total prediction           & $22.1 \pm 3.8$  & $17.9 \pm 3.5$
\\
  \hline
  Data                             & 24             & 20    \\
  \hline
  \end{tabular}
  \label{tab:evt}
\end{table}

The measured yield of data events is well described by the theoretical predictions, which include the EW contribution.
A CL$_\mathrm{s}$ based method~\cite{ATLAS:2011tau,0954-3899-28-10-313,Junk1999435} is used to estimate the upper limit on the EW
signal strength $\mu_{\text{sig}}$, which is defined as the ratio of the measured to the expected signal yield.
Combining four $m_{jj}$ bins from the two decay channels gives an upper limit of 4.3 times the SM
EW prediction at a 95\% confidence level (CL), compared to an expected limit of 2.0 from the background-only hypothesis.

The measured signal strength can be translated into the fiducial cross section $\sigma_{\text{fid}}$ using
the generated cross sections of the simulated samples $\sigma_{\text{gen}}$ and an acceptance $\epsilon_{\text{acc}}$ for the total cross section from the fiducial region to the signal region:
$\sigma_{\text{fid}}=\sigma_{\text{gen}}\mu_{\mathrm{sig}}\epsilon_{\text{acc}}$.
The fiducial cross section is reported in a region defined as follows:

\begin{itemize}
\item $\pt^{j1}>30\GeV$, $\abs{\eta^{j1}}<4.7$;
\item $\pt^{j2}>30\GeV$, $\abs{\eta^{j2}}<4.7$;
\item $m_{jj}>700\GeV$, $\abs{\Delta\eta(j,j)}> 2.4$;
\item $\pt^{\ell}>20\GeV$, $\abs{\eta^{\ell}}< 2.4$;
\item $\pt^{\gamma}>20\GeV$, $\abs{\eta^{\gamma}}< 1.4442$;
\item $\abs{\ptvecmiss}>20\GeV$;
\item $\Delta R_{jj}, \Delta R_{\ell j}, \Delta R_{\gamma j}, \Delta R_{\ell \gamma} >  0.4$.
\end{itemize}

This phase space corresponds to the acceptance of the CMS detector, with a minimal number of
additional selections on $m_{jj}$ and $\abs{\Delta\eta(j,j)}$ to ensure that the VBS contribution is large.
It does not include requirements on the Zeppenfeld variable and the $|\Delta\phi_{\PW\gamma,jj}|$ variable, which are applied at the reconstruction level.
The acceptance corrections for these selections are $0.289 \pm 0.001$ for the EW cross section and $0.174 \pm 0.002$ for the QCD one, where
we include both PDF and scale uncertainties.

The measured cross sections and signal strengths are summarized in Table~\ref{tab:qcdewkXS}, and the measured results are in good agreement with the theoretical predictions.
The EW signal strength is measured to be $\hat{\mu}_{\text{sig}} = 1.78^{+0.99}_{-0.76}$.
Considering both the EW and QCD contributions as a signal, the signal strength is measured to be $0.99^{+0.21}_{-0.19}$.
The EW fraction is found to be 27.1\% in the search region and 25.8\% in the fiducial region.
The significances for both cases are also determined: for the EW signal, the observed (expected) significance is found to be
2.7 (1.5) standard deviations; for the EW+QCD signal, it is found to be 7.7 (7.5) standard deviations.
The measured cross section in the fiducial region is
$10.8 \pm 4.1 \stat \pm 3.4 \syst \pm 0.3 \lum\unit{fb}$
for the EW-induced $\PW\gamma$+2\,jets production
and $23.2 \pm 4.3 \stat \pm 1.7 \syst \pm 0.6\lum\unit{fb}$ for the total $\PW\gamma$+2\,jets production.

\begin{table}[tbht!]
\centering
  \topcaption{Summary of the measured and predicted observables. }
  \label{tab:qcdewkXS}
\resizebox{\textwidth}{!}{
  \begin{tabular} {l l l}
  \hline
  Items    &    EW measurement      &    EW+QCD measurement       \\
  \hline
     Signal strength $\hat{\mu}_{\text{sig}}$               &    1.78$^{+0.99}_{-0.76}$  & 0.99$^{+0.21}_{-0.19}$      \\
   Observed (expected) significance &    2.7 (1.5) standard deviations    &  7.7 (7.5) standard deviations        \\
   Theoretical cross section (fb)   &$ 6.1 \pm1.2\,\text{(scale)}\pm 0.2\,\mathrm{(PDF)}$                        &   $23.5\pm5.3\,\text{(scale)}\pm0.8\,\mathrm{(PDF)}$  \\
   Measured cross section (fb) & $10.8 \pm 4.1 \stat \pm 3.4 \syst \pm 0.3 \lum$   &   $23.2 \pm 4.3 \stat \pm 1.7 \syst \pm 0.6 \lum$ \\
  \hline
  \end{tabular}
}
\end{table}

\section{Limits on anomalous quartic gauge couplings}
\label{sec:limits_pT}

\begin{figure}[h]
  \begin{center}
    \includegraphics[width=0.75\textwidth]{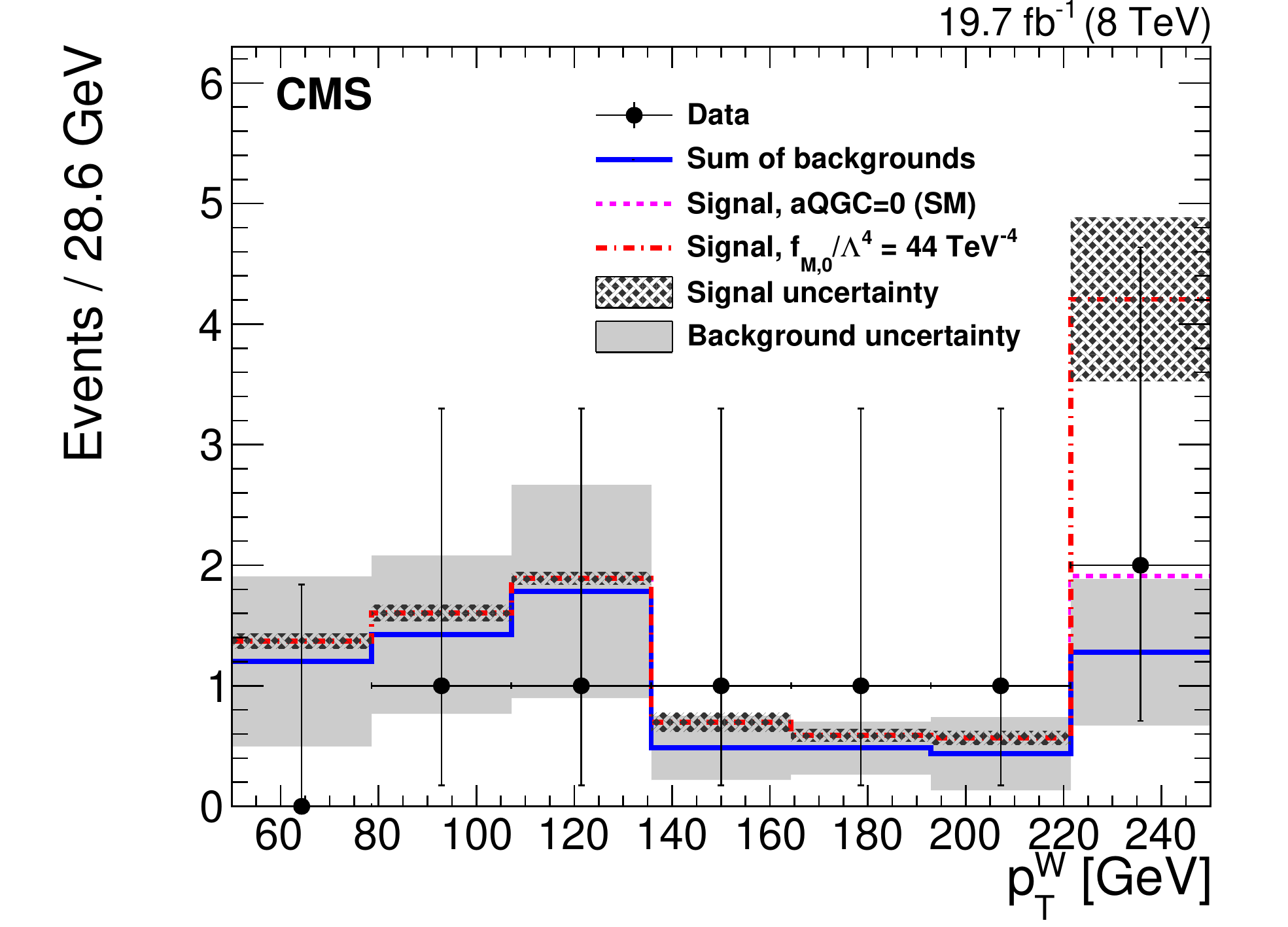} \\
    \caption{ Comparison of predicted and observed $\pt^{\PW}$ distributions with the combined electron and muon channels. The last $\pt^{\PW}$ bin has been extended to include the overflow contribution. The dash-dotted line depicts a representative signal distribution with anomalous coupling parameter $f_{M,0}/\Lambda^{4}=44\TeV^{-4}$ and the dashed line shows the same distribution corresponding to the SM case.
The bands represent the statistical and systematic uncertainties in signal and background predictions summed in quadrature. The data are shown with statistical uncertainties only.
    }
    \label{fig:wpt_shape}
  \end{center}
\end{figure}

Following Ref.~\cite{Eboli:2006wa}, we parameterize the aQGCs in a formalism that maintains $SU(2)_L \otimes U(1)_Y$ gauge
symmetry and leads to 14 possible dimension-eight operators that contribute to the signal.
The $\mathcal{L}_{M,5}$ operator is found to be non-Hermitian and needs to be replaced by a summation of the original and its Hermitian conjugate (see Appendix~\ref{sec:aQGCth} for the definition).
The presence of aQGCs should lead to an enhancement of the EW $\PW\gamma$+2\,jets cross section, which should become more pronounced at the high-energy tails of some distributions.
As shown in Fig.~\ref{fig:wpt_shape}, the $\pt^{\PW}$ distribution is sensitive to
the aQGCs and therefore is used to set limits. We choose a $\pt^{\PW}$ distribution binned over the range
50--250\GeV, with the overflow contribution included in the last bin.
The shape of the distribution at high $\pt^{\PW}$ is used to extract aQGC
limits. These limits are not sensitive to small variations in the
number of bins or range used for the $\pt^{\PW}$ distribution.
The events are selected with the baseline selections from Section~\ref{sec:recosel}, with the following additional
requirements:
$\abs{y_{\PW\gamma} -(y_{j1} + y_{j2})/2} < 1.2$, $\abs{\Delta\eta(j1,j2)}>2.4$, and
$\pt^{\gamma} > 200\GeV$.
A tight $\pt^{\gamma}$ selection is applied to reach higher expected significance for the possible aQGC signal in the EW $\PW\gamma$+2\,jets process.

The stringent selections above lead to increased statistical uncertainties in the estimations of the backgrounds. The second
largest uncertainty comes from the scale variations in the predicted aQGC signal. Other uncertainties include the signal PDF
choice, integrated luminosity, trigger efficiency, and lepton and photon efficiencies.

The search is performed for each aQGC parameter separately, while setting all other parameters to their SM values.
Each signal sample, representing a
different aQGC prediction, is generated at LO using the reweight method in \MADGRAPH~\cite{Alwall:2014hca}. For
each aQGC case, we compute the aQGC/SM event yield ratios for all $\pt^{\PW}$ bins from this sample and use these ratios to
rescale the SM signal shape to the enhanced aQGC shape. Then we consider the following test statistic:
\begin{equation}
t_{\alpha} = -2 \ln\dfrac{\mathcal{L}(\alpha,{\hat{\hat{\boldsymbol{\theta}}}})}{\mathcal{L}(\hat{\alpha},\hat{\boldsymbol{\theta}})},
\end{equation}
where the likelihood function is constructed in two lepton channels and then combined for the calculation.
The $\alpha$ term represents the aQGC point being tested, and $\boldsymbol{\theta}$ the nuisance parameters.
The $\hat{\hat{\boldsymbol{\theta}}}$ nuisance parameters correspond to the maximum of the
likelihood at the point $\alpha$, while $\hat{\alpha}$ and $ \hat{\boldsymbol{\theta}}$ correspond to the global
maximum of the likelihood.
This test statistic is assumed to follow a $\chi^2$ distribution~\cite{1943Wald,wilks1938}. One can therefore extract the
limits directly by using the delta log-likelihood function $\Delta \mathrm{NLL} = t_{\alpha}/2$~\cite{Khachatryan:2014jba}.
Table~\ref{tab:limit_noff} lists 95\% CL exclusion limits for all parameters.

\begin{table}[htb]
\centering
  \topcaption{Observed and expected shape-based exclusion limits for the aQGC parameters at 95\% CL, without any form factors.}
  \label{tab:limit_noff}
\scalebox{1.0}{
  \begin{tabular}{l l}
  \hline
  Observed limits ($\TeV^{-4}$) & Expected limits ($\TeV^{-4}$) \\
  \hline
   $-77 < f_{M,0}/\Lambda^{4} < 74$   &   $-47 < f_{M,0}/\Lambda^{4} < 44$  \\
   $-125 < f_{M,1}/\Lambda^{4} < 129$ &   $-72 < f_{M,1}/\Lambda^{4} < 79$  \\
   $-26 < f_{M,2}/\Lambda^{4} < 26$   &   $-16 < f_{M,2}/\Lambda^{4} < 15$  \\
   $-43 < f_{M,3}/\Lambda^{4} < 44$   &   $-25 < f_{M,3}/\Lambda^{4} < 27$  \\
   $-40 < f_{M,4}/\Lambda^{4} < 40$   &   $-23 < f_{M,4}/\Lambda^{4} < 24$  \\
   $-65 < f_{M,5}/\Lambda^{4} < 65$   &   $-39 < f_{M,5}/\Lambda^{4} < 39$  \\
   $-129 < f_{M,6}/\Lambda^{4} < 129$   &   $-77 < f_{M,6}/\Lambda^{4} < 77$ \\
   $-164 < f_{M,7}/\Lambda^{4} < 162$   &   $-99 < f_{M,7}/\Lambda^{4} < 97$ \\
   $-5.4 < f_{T,0}/\Lambda^{4} < 5.6$   &   $-3.2 < f_{T,0}/\Lambda^{4} < 3.4$ \\
   $-3.7 < f_{T,1}/\Lambda^{4} < 4.0$   &   $-2.2 < f_{T,1}/\Lambda^{4} < 2.5$ \\
   $-11 < f_{T,2}/\Lambda^{4} < 12$   &   $-6.3 < f_{T,2}/\Lambda^{4} < 7.9$  \\
   $-3.8 < f_{T,5}/\Lambda^{4} < 3.8$   &   $-2.3 < f_{T,5}/\Lambda^{4} < 2.4$  \\
   $-2.8 < f_{T,6}/\Lambda^{4} < 3.0$   &   $-1.7 < f_{T,6}/\Lambda^{4} < 1.9$  \\
   $-7.3 < f_{T,7}/\Lambda^{4} < 7.7$   &   $-4.4 < f_{T,7}/\Lambda^{4} < 4.7$  \\
  \hline
  \end{tabular}}
\end{table}

\begin{figure*}[bhtp]
\centering
   \includegraphics[width=0.95\textwidth]{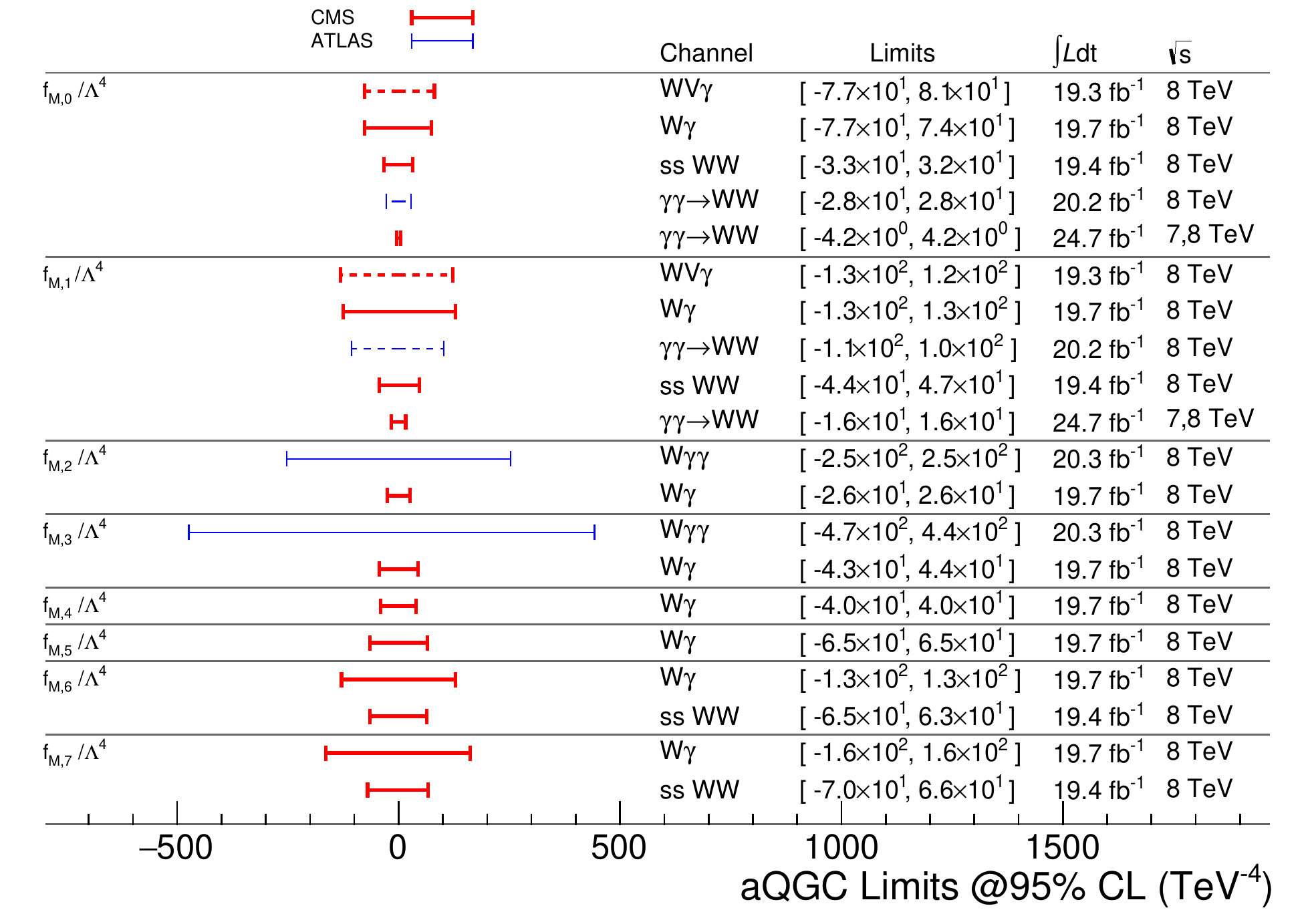} \\
   \includegraphics[width=0.95\textwidth]{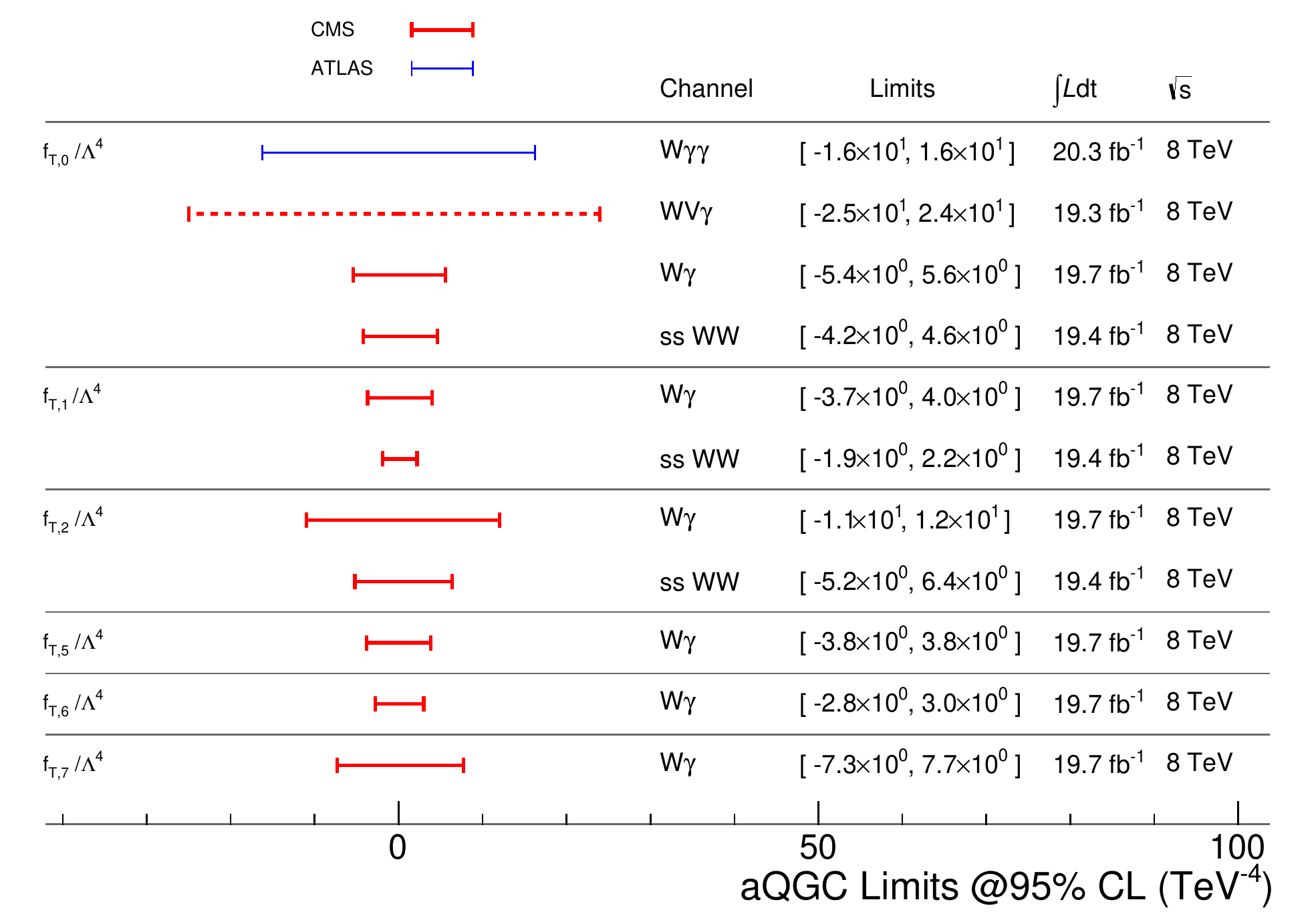}
   \caption{ Comparison of the limits on the dimension-eight aQGC parameters obtained
from this study $\PW\gamma$, together with results from the production of WV$\gamma$~\cite{Chatrchyan:2014bza}, same-sign WW~\cite{Khachatryan:2014sta}, exclusive $\gamma\gamma\to\PW\PW$ in ATLAS and CMS~\cite{Aaboud:2016dkv,Chatrchyan:2013foa,Khachatryan:2016mud}, and $\PW\gamma\gamma$ in ATLAS~\cite{Aad:2015uqa}. The limits from the CMS experiment are represented by thicker lines. The limits that are translated from another formalism are represented with dashed lines; details are found in Ref.~\cite{Chatrchyan:2014bza}.
}
\label{fig:WWAAcomparison}
\end{figure*}

Because of the nonrenormalizable nature of higher-dimensional operators, any nonzero aQGC parameter violates unitarity at high energies.
An effective theory is therefore only valid at low energies, and we need to check that the energy scale we probe is less
than a new physics scale and does not violate unitarity.
Sometimes a form factor is introduced to unitarize the high-energy contribution within that energy range; however, the form factor complicates the limit-setting
procedure and makes it difficult to compare results among experiments.
We use \textsc{vbfnlo} without any form factors to
calculate the unitarity bound corresponding to the maximum aQGC enhancements, which would conserve unitarity within the range of energies probed at the 8\TeV LHC~\cite{Baglio:2014uba,Gounaris:1993fh}.
We find that unitarity is violated in many cases.
We compare our results, in a consistent way, with existing limits on aQGC parameters in Fig.~\ref{fig:WWAAcomparison}, where the aQGC convention used in \textsc{vbfnlo} has been transformed to the one that is used in our analysis.
Existing competitive limits
include the results from WV$\gamma$ production~\cite{Chatrchyan:2014bza}, same-sign WW production~\cite{Khachatryan:2014sta}, exclusive $\gamma\gamma\to\PW\PW$ production at the ATLAS and the CMS experiments~\cite{Aaboud:2016dkv,Chatrchyan:2013foa,Khachatryan:2016mud}, and $\PW\gamma\gamma$ production at the ATLAS experiment~\cite{Aad:2015uqa}.
The limits on the
$a_0^W/\Lambda^2$ and $a_C^W/\Lambda^2$ couplings in these references are transformed to ours by using Eq.~(2) in
Ref.~\cite{Chatrchyan:2014bza}, with the constraint of $f_{M,0}/\Lambda^4 = 2  f_{M,2}/\Lambda^4$ and
$f_{M,1}/\Lambda^4 = 2  f_{M,3}/\Lambda^4$. All of the aQGC limits shown are calculated without a
form factor.

\section{Summary}
\label{sec:summary}

A search for EW-induced $\PW\gamma$+2\,jets production and aQGCs has been presented based on events containing a W boson that decays to a lepton and a neutrino, a hard photon, and two jets with large pseudorapidity separation. The data analyzed correspond to an integrated
luminosity of 19.7\fbinv collected in proton-proton collisions at $\sqrt{s} = 8\TeV$ with the CMS detector at the LHC.
An excess is observed above the expectation from QCD-induced $\PW\gamma$+2\,jets and other backgrounds, with an observed (expected) significance of 2.7 (1.5) standard deviations.
The corresponding cross section within the VBS-like fiducial region is measured to be
$10.8 \pm 4.1 \stat \pm 3.4 \syst \pm 0.3 \lum \unit{fb}$,
 which is consistent with the SM prediction of EW-induced signal. In the same fiducial region, the total cross section for $\PW\gamma$+2\,jets is measured to be
$23.2 \pm 4.3 \stat \pm 1.7 \syst \pm 0.6 \lum \unit{fb}$, which is consistent with the SM EW+QCD prediction.
Exclusion limits for aQGC parameters $f_{M,0-7}/\Lambda^4$, $f_{T,0-2}/\Lambda^4$, and $f_{T,5-7}/\Lambda^4$ are set at
95\% CL\@. Competitive limits are obtained for several parameters and first limits are set on the
$f_{M,4}/\Lambda^4$ and $f_{T,5-7}/\Lambda^4$ parameters.

\section*{Acknowledgements}

We congratulate our colleagues in the CERN accelerator departments for the excellent performance of the LHC and thank the
technical and administrative staffs at CERN and at other CMS institutes for their contributions to the success of the CMS
effort. In addition, we gratefully acknowledge the computing centers and personnel of the Worldwide LHC Computing Grid for
delivering so effectively the computing infrastructure essential to our analyses. Finally, we acknowledge the enduring support
for the construction and operation of the LHC and the CMS detector provided by the following funding agencies: BMWF and FWF
(Austria); FNRS and FWO (Belgium); CNPq, CAPES, FAPERJ, and FAPESP (Brazil); MES (Bulgaria); CERN; CAS, MoST, and NSFC (China);
COLCIENCIAS (Colombia); MSES and CSF (Croatia); RPF (Cyprus); MoER, SF0690030s09, and ERDF (Estonia); Academy of Finland, MEC,
and HIP (Finland); CEA and CNRS/IN2P3 (France); BMBF, DFG, and HGF (Germany); GSRT (Greece); OTKA and NIH (Hungary); DAE and
DST (India); IPM (Iran); SFI (Ireland); INFN (Italy); NRF and WCU (Republic of Korea); LAS (Lithuania); MOE and UM (Malaysia);
CINVESTAV, CONACYT, SEP, and UASLP-FAI (Mexico); MBIE (New Zealand); PAEC (Pakistan); MSHE and NSC (Poland); FCT (Portugal);
JINR (Dubna); MON, RosAtom, RAS, and RFBR (Russia); MESTD (Serbia); SEIDI and CPAN (Spain); Swiss Funding Agencies
(Switzerland); NSC (Taipei); ThEPCenter, IPST, STAR, and NSTDA (Thailand); TUBITAK and TAEK (Turkey); NASU (Ukraine); STFC
(United Kingdom); and DOE and NSF (USA).

Individuals have received support from the Marie-Curie program and the European Research Council and EPLANET (European
Union); the Leventis Foundation; the A. P. Sloan Foundation; the Alexander von Humboldt Foundation; the Belgian Federal Science
Policy Office; the Fonds pour la Formation \`a la Recherche dans l'Industrie et dans l'Agriculture (FRIA-Belgium); the
Agentschap voor Innovatie door Wetenschap en Technologie (IWT-Belgium); the Ministry of Education, Youth and Sports (MEYS) of
Czech Republic; the Council of Science and Industrial Research, India; the Compagnia di San Paolo (Torino); the HOMING PLUS
programme of Foundation for Polish Science, cofinanced by EU, Regional Development Fund; and the Thalis and Aristeia programmes
cofinanced by EU-ESF and the Greek NSRF.

\appendix
\section{Anomalous quartic gauge coupling parameterization}
\label{sec:aQGCth}

Gauge boson self-interactions are fixed by the gauge symmetries of the SM. To investigate possible deviations from the SM, we
parameterize the aQGCs in a formalism that maintains the $SU(2)_L \otimes U(1)_Y$ gauge symmetry. As a natural extension to the SM,
the lowest order pure anomalous quartic couplings arise from dimension-eight operators. This analysis adopts the following
effective Lagrangian containing such aQGCs~\cite{Eboli:2006wa}:

\begin{eqnarray}\label{lagrangian}
\mathcal{L}_{\mathrm{aQGC}} &=&
 \dfrac{f_{M,0}}{\Lambda^{4}}\;\mbox{Tr}\left[ \mathbf{W}_{\mu\nu} \mathbf{W}^{\mu\nu} \right] \times \left[  (D_{\beta}\Phi)^{\dagger} D^{\beta}\Phi \right] +
 \dfrac{f_{M,1}}{\Lambda^{4}}\;\mbox{Tr}\left[ \mathbf{W}_{\mu\nu} \mathbf{W}^{\nu\beta}\right] \times \left[  (D_{\beta}\Phi)^{\dagger} D^{\mu}\Phi \right] \nonumber \\
 &+& \dfrac{f_{M,2}}{\Lambda^{4}}\;\left[ B_{\mu\nu} B^{\mu\nu}\right] \times \left[  (D_{\beta}\Phi)^{\dagger} D^{\beta}\Phi \right] +
 \dfrac{f_{M,3}}{\Lambda^{4}}\;\left[ B_{\mu\nu} B^{\nu\beta}\right] \times \left[  (D_{\beta}\Phi)^{\dagger} D^{\mu}\Phi \right] \nonumber \\
 &+& \dfrac{f_{M,4}}{\Lambda^{4}}\;\left[  (D_{\mu}\Phi)^{\dagger} \mathbf{W}_{\beta\nu}  D^{\mu}\Phi \right] \times B^{\beta\nu} +
\dfrac{f_{M,5}}{\Lambda^{4}}\; \times \dfrac{1}{2} \left[ (D_{\mu}\Phi)^{\dagger} \mathbf{W}_{\beta\nu}  D^{\nu}\Phi + (D^{\nu}\Phi)^{\dagger} \mathbf{W}_{\beta\nu}  D_{\mu}\Phi \right] \times B^{\beta\mu} \nonumber \\
 &+& \dfrac{f_{M,6}}{\Lambda^{4}}\;\left[  (D_{\mu}\Phi)^{\dagger} \mathbf{W}_{\beta\nu} \mathbf{W}^{\beta\nu} D^{\mu}\Phi \right] +
\dfrac{f_{M,7}}{\Lambda^{4}}\;\left[  (D_{\mu}\Phi)^{\dagger} \mathbf{W}_{\beta\nu} \mathbf{W}^{\beta\mu} D^{\nu}\Phi \right] \nonumber \\
 &+& \dfrac{f_{T,0}}{\Lambda^4} Tr[\mathbf{W}_{\mu \nu} \mathbf{W}^{\mu \nu}] \times Tr[\mathbf{W}_{\alpha \beta} \mathbf{W}^{\alpha \beta}] +
\dfrac{f_{T,1}}{\Lambda^4} Tr[\mathbf{W}_{\alpha \nu} \mathbf{W}^{\mu \beta}] \times Tr[\mathbf{W}_{\mu \beta} \mathbf{W}^{\alpha \nu}] \nonumber \\
 &+& \dfrac{f_{T,2}}{\Lambda^4} Tr[\mathbf{W}_{\alpha \mu} \mathbf{W}^{\mu \beta}] \times Tr[\mathbf{W}_{\beta \nu} \mathbf{W}^{\nu \alpha}] +
\dfrac{f_{T,5}}{\Lambda^4} Tr[\mathbf{W}_{\mu \nu} \mathbf{W}^{\mu \nu}] \times B_{\alpha \beta} B^{\alpha \beta} \nonumber \\
 &+& \dfrac{f_{T,6}}{\Lambda^4} Tr[\mathbf{W}_{\alpha \nu} \mathbf{W}^{\mu \beta}] \times B_{\mu \beta} B^{\alpha \nu} +
\dfrac{f_{T,7}}{\Lambda^4} Tr[\mathbf{W}_{\alpha \mu} \mathbf{W}^{\mu \beta}] \times B_{\beta \nu} B^{\nu \alpha},
\end{eqnarray}

where $\Phi$ represents the Higgs doublet, $B_{\mu \nu}$ and $W_{\mu \nu}^i$ are the associated field strength tensors of the
$U(1)_Y$ and $SU(2)_L$ gauge symmetries, and $\mathbf{W}_{\mu \nu} \equiv \sum_j W_{\mu \nu}^j \sigma^j/2$. The
$f_{T}/\Lambda^{4}$ associated operators characterize the effect of new physics on the scattering of transversely
polarized vector bosons, and $f_{M}/\Lambda^{4}$ includes mixed transverse and longitudinal scatterings; however,
pure longitudinal scattering effects do not occur in the $\PW\gamma$ final state due to the presence of the photon. The listed
operators include all contributions to the $\PW\PW\gamma\gamma$ and $\PW\PW\Z\gamma$ vertices. In this paper, we set $c=1$ to describe
energy, momentum, and mass in units of GeV.

Any nonzero value in aQGCs will lead to tree-level unitarity violation at sufficiently high energy and could be unitarized
with a suitable form factor; however the unitarization depends on the detailed structure of new physics, which is not known a
priori. Following Ref.~\cite{Chatrchyan:2014bza}, the choice is made to set limits without using a form factor.

\bibliography{auto_generated}

\cleardoublepage \section{The CMS Collaboration \label{app:collab}}\begin{sloppypar}\hyphenpenalty=5000\widowpenalty=500\clubpenalty=5000\textbf{Yerevan Physics Institute,  Yerevan,  Armenia}\\*[0pt]
V.~Khachatryan, A.M.~Sirunyan, A.~Tumasyan
\vskip\cmsinstskip
\textbf{Institut f\"{u}r Hochenergiephysik,  Wien,  Austria}\\*[0pt]
W.~Adam, E.~Asilar, T.~Bergauer, J.~Brandstetter, E.~Brondolin, M.~Dragicevic, J.~Er\"{o}, M.~Flechl, M.~Friedl, R.~Fr\"{u}hwirth\cmsAuthorMark{1}, V.M.~Ghete, C.~Hartl, N.~H\"{o}rmann, J.~Hrubec, M.~Jeitler\cmsAuthorMark{1}, A.~K\"{o}nig, I.~Kr\"{a}tschmer, D.~Liko, T.~Matsushita, I.~Mikulec, D.~Rabady, N.~Rad, B.~Rahbaran, H.~Rohringer, J.~Schieck\cmsAuthorMark{1}, J.~Strauss, W.~Treberer-Treberspurg, W.~Waltenberger, C.-E.~Wulz\cmsAuthorMark{1}
% ...
\vskip\cmsinstskip
% ...
1:~~Also at Vienna University of Technology, Vienna, Austria\\
% ...

\end{sloppypar}
\end{document}